\documentclass[11pt]{article}
\usepackage{moriond,epsfig}

\def\nn{\nonumber}

\def\be{\begin{equation}}
\def\ee{\end{equation}}
\def\bea{\begin{eqnarray}}
\def\eea{\end{eqnarray}}

\begin{document}
\vspace*{4cm}
\title{GRAND UNIFICATION IN EXTRA DIMENSIONS AND PROTON DECAY}

\author{FERRUCCIO FERUGLIO}

\address{Department of Physics, University of Padova and 
I.N.F.N, Sezione di Padova, Via Marzolo 8\\
Padova - Italy}

\maketitle\abstracts{We discuss baryon and lepton violation 
in the context of a 
simple 5-dimensional grand unified model, based on the 
orbifold $S^1/(Z_2\times Z_2')$. While gauge and Higgs
degrees of freedom live in the bulk, matter is located on
the boundaries of the space-time.
We show
that proton decay is naturally suppressed or even forbidden by 
suitable implementations of the parity symmetries in the matter sector. 
The corresponding mechanism does not affect the SU(5) description of fermion
masses also including neutrinos.}

The idea that strong and electroweak interactions may possess
a common description in the framework of a grand unified theory
(GUT) is very attractive. Supersymmetric GUTs, with superpartners
of the ordinary particles at the TeV scale as required by
the solution of the hierarchy problem, predict a successful
gauge couplings unification \cite{dg} at a very large mass scale,
close to the gravitational scale and possibly coinciding with it \cite{hw}. 
Moreover, 
the violation of the lepton number in the vicinity of the GUT scale
would provide an elegant description of the observed smallness of 
neutrino masses. One of the distinctive features of GUTs,
the violation of baryon number (B) is also one of the necessary conditions
to generate the baryon asymmetry of the universe starting from symmetric 
conditions. 

Despite their beauty, GUTs suffer from several difficulties that
render rather cumbersome their specific realization in the context of 
conventional, four dimensional models. Perhaps the most serious 
problem is represented by the doublet-triplet (DT) splitting that
in realistic GUTs can only be achieved at the cost of a quite 
complicated Higgs sector. Furthermore, minimal GUTs
predict a proton lifetime that, although affected by large theoretical
uncertainties, is on the verge of being experimentally excluded \cite{sk}.
These drawbacks provide a strong motivation to look for alternative 
formulations of GUTs, also by going beyond the conventional framework.
 
It has been recently observed \cite{kawa} that the DT splitting problem
can find an economic and elegant solution if the GUT gauge symmetry
is realized in 5 (or more) space-time dimensions and broken down to the 
Standard Model (SM) gauge symmetry by the compactification of the extra
dimension(s). 
Of course the idea that extra dimensions may offer
a natural framework for grand unification is an old one.
Already in the late seventies, extended gauge 
symmetries were found by building supergravity theories in higher 
dimensions \cite{cj} and, subsequently, by looking for non-anomalous 
supergravity/superstring theories \cite{gs}.
Furthermore it was soon suggested that in models with extra dimensions 
the grand unification scale could be set by the inverse 
compactification radius \cite{fayet}. 
It was also clear that the compactification
process could offer new ways of breaking the gauge symmetry,
in particular with the help of singular manifolds \cite{orbi}.
Indeed the model of ref. \cite{kawa} consists of a 5-dimensional N=2 supersymmetric
GUT where the compactification of the fifth dimension on $S^1/(Z_2\times Z_2')$
breaks at the same time N=2 down to N=1 and SU(5) down to 
SU(3)$\times$SU(2)$\times$U(1). The novelty of this model is the specific
mechanism employed to obtain the DT splitting.

\begin{figure}[t]
\centerline{\psfig{figure=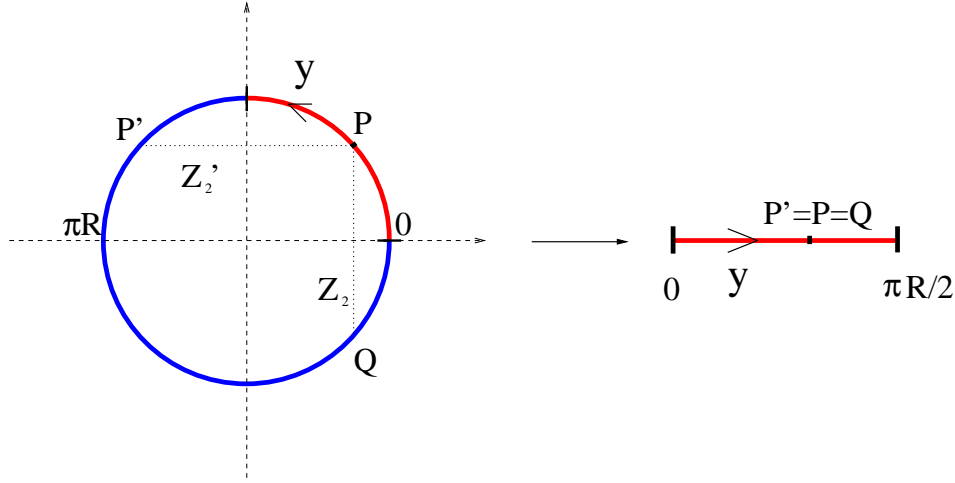,height=2.5in}}
\caption{Orbifold $S^1/(Z_2\times Z_2')$. The points $y=0$ and $y=\pi R$ 
($y=\pm \pi R/2$) are fixed under $Z_2$ ($Z_2'$) and related under $Z_2'$ 
($Z_2$).
\label{fig:orb}}
\end{figure}

The fifth dimension is spanned by a coordinate $y$ parameterizing
a circle $S^1$ with the identification of points related by the
discrete symmetries $Z_2$ and $Z_2'$ \cite{bar}. These
are reflections symmetries about orthogonal diameters of
the circle (see fig.1): 
$y\to -y$ and $y' \to -y'$, respectively ($y'=y-\pi R/2$).
The resulting orbifold $S^1/(Z_2\times Z_2')$ can be thought as the
arc going from $y=0$ to $y=\pi R/2$.
The space-time has a 5-dimensional bulk, $0 < y < \pi R/2$,
and two 4-dimensional boundaries at $y=0$ and $y=\pi R/2$.
The metric is everywhere flat. The generic bulk field $\phi(x,y)$, depending
on all 5-dimensional coordinates, has well-defined
($P$, $P'$) $(P,P'=\pm 1)$ parities under ($Z_2$, $Z_2'$).
There are only four possible cases: $\phi_{++}$, $\phi_{+-}$,
$\phi_{-+}$ and $\phi_{--}$, whose Fourier expansions give
rise to 4-dimensional modes of masses $2n/R$, $(2n+1)/R$, $(2n+1)/R$
and $(2n+2)/R$, ($n\ge 0$), respectively. Only the bulk field of the 
type $\phi_{++}$ has a zero mode.

\begin{table}[b]
\caption{Parity assignment and masses ($n\ge 0$) for
gauge vector bosons and Higgs supermultiplets. \label{tab1}}
\vspace{0.4cm}
\begin{center}
\begin{tabular}{|c|c|c|}   
\hline
& & \\                         
$(P,P')$ & field & mass\\ 
& & \\
\hline
& & \\
$(+,+)$ &  $A^a_{\mu}$, $H^D_u$, $H^D_d$ & $\frac{2n}{R}$\\
& & \\
\hline
& & \\
$(+,-)$ &  $A^{\hat{a}}_{\mu}$, $H^T_u$, $H^T_d$ & $\frac{(2n+1)}{R}$ \\
& & \\ 
\hline
\end{tabular}
\end{center}
\end{table}
The theory contains a bulk vector supermultiplet that includes a 
set of gauge bosons $A^A_\mu$ ($\mu=0,...,3$) ($A=1,...24$)
together with their 5-dimensional completions ($\mu=5$) and their 
supersymmetric
partners. The index $A$ will be denoted by $a$ when referring
to SU(3)$\times$SU(2)$\times$U(1) and by ${\hat a}$ when 
indicating the coset SU(5)/SU(3)$\times$SU(2)$\times$U(1).
In the Higgs sector there are bulk N=1 chiral multiplets 
$H_u$ and $H_d$ transforming as 5 and $\bar 5$ under SU(5) and 
belonging to distinct hypermultiplets of N=2 supersymmetry. They contain SU(2) 
doublets $H^D_{u,d}$ and color triplets $H^T_{u,d}$.
The $(Z_2,Z_2')$ parities of the relevant bulk fields are shown in
table 1. The only massless vector bosons of the theory are the zero modes
of $A^a_\mu$, which are identified with the gluons and the electroweak
gauge bosons. The other vector bosons have masses of order $1/R$: SU(5) is broken down to SU(3)$\times$SU(2)$\times$U(1) (see fig. 2). 
The length of the radius $R$ is of order $(10^{16}~GeV)^{-1}$.
In $H_{u,d}$, color triplets are automatically splitted from SU(2) doublets, 
since the only massless scalars are the zero modes of $H^D_{u,d}$, 
while the remaining modes have masses of $O(1/R)$.
The 5-dimensional parameters of gauge transformations,
$\alpha^a(x,y)$ and $\alpha^{\hat a}(x,y)$ have $(P,P')$ parities
equal to $(+,+)$ and $(+,-)$, respectively. This means 
that in $y=\pi R/2$ only $\alpha^a$ is non-vanishing
and the transformations reduce to those of a (5-dimensional)
SU(3)$\times$SU(2)$\times$U(1) group. On the boundary at $y=0$
fields feel both $\alpha^a$ and $\alpha^{\hat a}$ parameters.
\begin{figure}[t]
\centerline{\psfig{figure=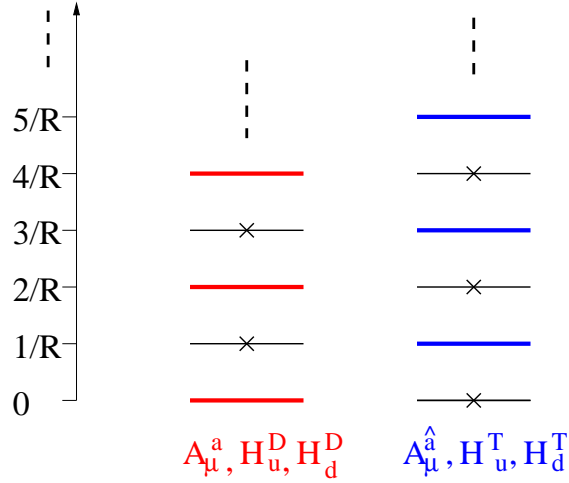,height=2.5in}}
\caption{Modes in the Fourier expansion of the bulk fields. Crossed
levels are eliminated by the orbifold projection thus breaking 4-dimensional
SU(5) multiplets into disjoint components.
\label{fig:lev}}
\end{figure}
Matter fields cannot be bulk fields and can only live 
on the boundaries \cite{af4}, either in $y=0$ or in $y=\pi R/2$. 
To motivate the 
introduction of matter in the SU(5)
representations 10 and $\bar 5$ (and N=1 chiral
multiplets), the natural choice is $y=0$, where the whole (5-dimensional)
gauge group is active.
To preserve the orbifold construction,  $10\equiv (Q,U^c,E^c)$
and ${\bar 5}\equiv (L,D^c)$ should be even under
$Z_2$, because $y=0$ is a fixed point under $Z_2$, and should possess
appropriate $Z_2'$ parities. The only $Z_2'$ parities that are
compatible with SU(5) are \cite{af4}:
\be
(Q,U^c,E^c)=\pm(+,-,-)~~~~~~~~~(L,D^c)=\pm(+,-)~~~. 
\label{par5}
\ee
None of these choices leads however to realistic 
SU(5)$\times Z_2\times Z_2'$ invariant Yukawa couplings,
if the coupling constants are independent from $y$ over $S^1$.
Indeed, the choice in eq. (\ref{par5}) implies,
for a single generation of matter fields,
that $10 10 H_u$ is odd under $Z_2'$. If such a term were
present in $y=0$ with a Yukawa coupling $y_u$, then the $Z_2'$ symmetry
would require the same term in $y=\pi R$ with the opposite
coupling $-y_u$. Thus the coupling of $10 10 H_u$, defined
on the whole circle $S^1$, would behave like a $Z_2'$ odd field.
Barring this interesting possibility \cite{hano}, 
$y_u$ should vanish and no mass for 
the up quark is obtained. Also with 3 generation no realistic spectrum
in the up sector can be recovered.

Realistic masses for matter fields are instead obtained from the
superpotential
\be
w=y_u~ Q U^c H_u^D + y_d~ Q D^c H_d^D + y_e~ L E^c H_d^D~~~,
\ee
provided each term is invariant under $(Z_2\times Z_2')$. 
If we bound ourselves to the case of Yukawa couplings
$y_u$, $y_d$ and $y_e$ constant over $S^1$, 
the orbifold symmetry implies that $Q$, $U^c$ and $D^c$ 
should have equal $Z_2'$ parities
and similarly for $L$ and $E^c$:
\be
(Q,U^c,E^c,L,D^c)=(P'_q,P'_q,P'_l,P'_l,P'_q)~~~~~~(P'_q,P'_l=\pm 1)~~~.
\label{par0}
\ee
The $Z_2\times Z_2'$ invariant Yukawa interactions can be defined
by first considering, at the boundary $y=0$, the superpotential
\be
w(y)=y_u~ 10~ 10~ H_u + y_d~ 10~ {\bar 5}~ H_d+...
\ee
where dots stand for R-parity violating terms. Given the decompositions
\be
10~ 10~ H_u= Q U^c H_u^D + \frac{1}{2} Q Q H_u^T + U^c E^c H_u^T~~~,\nn
\ee
\be
10~ {\bar 5}~ H_d= Q D^c H_d^D + L E^c H_d^D + Q L H_d^T + U^c D^c H_d^T~~~,\nn
\ee
and the $Z_2'$ parity assignments, we can separate in $w(y)$ an even part and an odd part 
under $Z_2'$: $w(y)=w_E(y)+w_O(y)$.
The odd part is projected out by requiring, at the $Z_2'$ mirror point $y=\pi R$, the
same interactions as in $y=0$:
\bea
w^{(4)}&=&\int dy \left[\delta(y)+\delta(-y+\pi R)\right] w(y)\nn\\
&=&\int dy \left[\delta(y)+\delta(-y+\pi R)\right] w_E(y)
\label{intyuk}
\eea
Eq. (\ref{intyuk}) is taken as the definition of 
$\left[\delta(y)+\delta(-y+\pi R)\right] w_E(y)$, the 5-dimensional Yukawa interaction.
The SU(5) gauge symmetry is thus 
explicitly violated by both the gauge and the Yukawa interactions 
at $y=0$. As a further consequence of the parity choice in (\ref{par0}), 
the B-violating terms
$\overline{\psi_Q} \bar\sigma^\mu T^{\hat a} \psi_{U^c} A^{\hat a}_\mu$
(where $\psi_M$ stands for the fermion member of the $M$ chiral multiplet
and $T^A$ are the SU(5) generators),
$QQ H^T_u$ and $U^c D^c H^T_d$ are odd under $Z_2'$ and vanish.
Hence the tree-level amplitudes from gauge boson or Higgsino exchange
contributing to proton decay also vanish. 
Dangerous combinations of
the dimension 4 operators $QLD^c$, $LLE^c$ and $U^cD^cD^c$
can be avoided by particular $Z_2'$ assignments like
\be
(Q,U^c,E^c,L,D^c)=(+,+,-,-,+)
\label{par}
\ee
Actually, with the choice of eq. (\ref{par}) the proton is stable.
Neutrino masses can be generated either by operators living on the boundaries
like $LH^D_dLH^D_d$ or by the presence of a $Z_2'$-odd, 
SU(5) singlet $\nu^c$. 
In the latter case both Dirac and Majorana
neutrino mass terms are allowed and the see-saw mechanism is
viable. A large mixing for atmospheric neutrinos can be driven
by a large mixing between right-handed $s$ and $b$ quarks via
the relation $y_e=y^T_d$. In conclusion, realistic fermion masses 
are obtained with parities that break SU(5) and that do not allow for 
tree-level proton decay amplitudes, via gauge boson or Higgsino
exchange. Suitable parities can make the proton stable, while allowing
for the desired terms that describe the observed neutrino oscillations.
 
The use of discrete symmetries to remove operators providing dangerous
contributions to the proton decay has been advocated long ago \cite{saya}.
The interesting feature of the model under discussion is that here such
symmetries are not introduced appositely to tame proton decay
but they are essential to the construction of the space-time orbifold
underlying the theory. The explicit SU(5) breaking due to the 
parity choice in (\ref{par}) is not welcome. It prevents a common
evolution of the gauge couplings after the unification scale.
It weakens the motivation for introducing matter in GUT representations.
However these unpleasant features might be peculiar of
the exploratory 5-dimensional model presented here and they
could hopefully be avoided in alternative constructions based on different 
orbifold symmetries or on higher space-time dimensions.
 
\section*{Acknowledgments}
I would like to thank Guido Altarelli for the enjoyable collaboration on which 
this talk is based. Many thanks go also to Andrea Brignole and  
Yasunori Nomura for useful discussions.
This work was partially supported by the European Program HPRN-CT-2000-00148 
(network Across The Energy Frontier).

\end{document}